\documentclass[prd,aps,twocolumn,floatfix]{revtex4}
\pdfoutput=1

\usepackage{psfrag,verbatim,bm}
\usepackage{hyperref}
\usepackage{amsmath,amssymb,graphicx}
\usepackage{epsfig}
\usepackage[usenames]{color}

\begin{document}

\title{Nonlinear Collapse in the Semilinear Wave Equation in AdS}

\author{Steven L. Liebling}
\affiliation{Department of Physics, Long Island University, New York 11548, USA}

\begin{abstract}
Previous studies of the semilinear wave equation in Minkowski space have shown
a type of critical behavior in which large initial data collapse to
singularity formation due to nonlinearities while small initial data
does not. Numerical solutions 
in spherically symmetric Anti-de Sitter~(AdS)
are  presented here which suggest
that, in contrast,  even small initial data collapse eventually.
Such behavior appears analogous to the recent result of Ref.~\cite{Bizon:2011gg}
that found that even weak, scalar initial data collapse gravitationally
to black hole
formation via a weakly turbulent instability.
Furthermore, the imposition of a reflecting boundary condition in the bulk
introduces a cut-off, below which initial data fails to collapse. This
threshold appears to arise because of the dispersion introduced by the boundary condition.
\end{abstract}

\maketitle

\noindent \emph{\bf Introduction:}
Recently, numerical studies of the gravitational collapse of a
scalar field in asymptotically Anti-de Sitter~(AdS) space found
that the scalar field collapses to a black hole eventually
for \emph{any} initial amplitude for generic initial
data~\cite{Bizon:2011gg,Jalmuzna:2011qw}.
This inevitability of black hole formation is understood as thermalization
within the corresponding conformal field theory~(CFT) on the boundary
of AdS according to AdS/CFT correspondence.
The nature of the boundary of AdS is such that it can be reached
in finite time and
therefore the bulk is a bounded domain. As noted in Ref.~\cite{Bizon:2011gg},
there is evidence for similar nonlinear behavior in other, non-gravitational
systems in bounded domains.
%that it is common for nonlinear
%systems in general (not just gravitational ones) in bounded domains
%to behave in a similar fashion.

It is in the context of such studies that one considers the dynamics of the
semilinear wave equation in AdS
for a complex scalar field $\phi$
with a nonlinear potential term 
%\begin{equation}
%\frac{\partial^2}{\partial t^2} \phi = \Delta \phi + \phi^p.
$\phi^p$
%\label{eq:wave}
%\end{equation}
where $p$ is an odd integer.
%by to the scalar wave equation.
Previous studies of this model (for a real scalar field)
in Minkowski space have shown that for 
$p=7$, the scalar field ``collapses'' to singularity formation for
large initial data and disperses for small initial
data~\cite{Liebling:2005ap,2004Nonli..17.2187B}. 
In AdS space, one might suspect that even small initial data,
which would otherwise disperse in Minkowski space, would instead
reflect off the AdS boundary repeatedly and eventually collapse.
Indeed, numerical solutions presented here suggest that such
behavior is present for the semilinear equation.  Furthermore,
placing a reflecting boundary at some finite radius places a lower
limit on the size of collapsing initial data, similar to that found
in the gravitating case~\cite{Buchel:2012uh}. Some speculation about
the nature of this change in behavior is given.\\

%%%%%%%%%%%%%%%%%%%%%%%%%%%%%%%%%%%%%%%%%%%%%%%%%
\begin{figure}
\vspace{-0.75in}
\centerline{\includegraphics[width=4in]{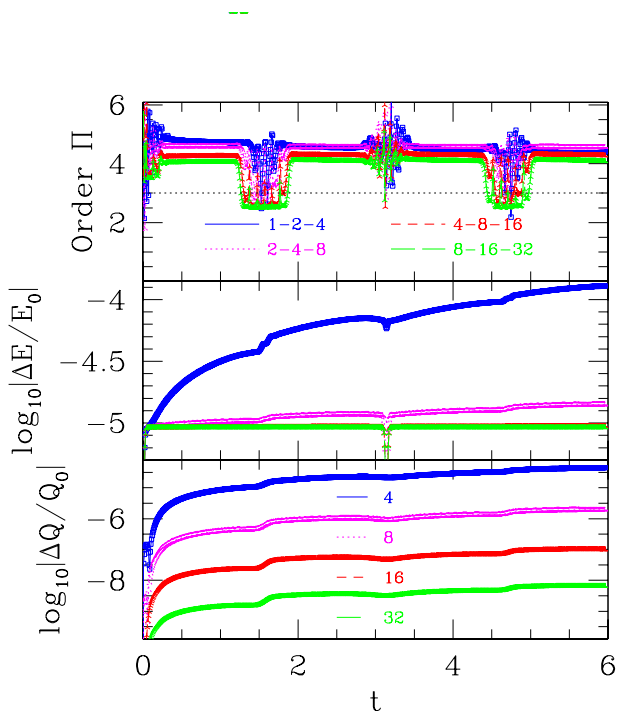}}
\vspace{-1.15in}
%\centerline{\includegraphics[width=4in]{convergence.ps}}
\caption{Demonstration of convergence through a bounce with $\epsilon=1$ and $\omega=5$.
Runs with successively doubled resolutions are compared where the run indicated as ``32'' has a resolution smaller by a factor of $2^{5}$ than the base resolution run ``1."(top) The order of
convergence for the field $\Pi_1$ as a function of time is shown. The other fields converge similarly. Runs were
carried out with multiples as shown of a base resolution. The plot shows that
the scheme is generally accurate roughly to fourth order, except when the pulse is
at a boundary where it drops to third order. (middle) The fractional change in energy
$\log_{10} \left|E(t)-E(0)\right| / E(0)$ as a function of time for the four highest-resolution
runs. The conservation of energy improves rapidly with resolution and conserves energy
to better than about one part in $10^5$. (bottom) The fractional change in charge
versus time. The conservation similarly improves with resolution.
}
\label{fig:convergence}
\end{figure}

\noindent \emph{\bf Implementation:}
Adopting essentially the same notation and form as in~\cite{Buchel:2012uh}, the metric is assumed to be that of spherically symmetric AdS
\begin{equation}
ds^2 = \frac{\ell^2}{\cos^2 x} \left(
                                     -dt^2
                                     +dx^2
                                     +\sin^2x \, d\Omega^2_{d-1}
                                        \right)  ,
\label{eq:metric}
\end{equation}
where $\ell$ is the scale-size of the AdS spacetime,
$d\Omega^2_{d-1}$ is the metric of $S^{d-1}$.
The domain extends from the origin, $x=0$, to the boundary of AdS, $x=\pi/2$.
Introducing the  auxiliary quantities
$\Phi_i \equiv \left(\partial/\partial x\right) \left( \phi_i \right)$
and
$\Pi_i \equiv \left(\partial/\partial t\right) \left( \phi_i \right)$,
one rescales according to
\begin{eqnarray}
\hat \phi_i & \equiv  & \frac{\phi_i}{\cos^{d-1}x} \, , \\
\hat \Pi_i  & \equiv  &                   \frac{\partial_t \phi_i}{\cos^{d-1}x} 
                  = \frac{\Pi_i}{\cos^{d-1}x} \, ,\\ 
\hat \Phi_i & \equiv  & \frac{\partial_x\phi_i}{\cos^{d-2}x}
                  = \frac{\Phi_i}{\cos^{d-2}x} \, .
\end{eqnarray}
Here, $\phi_i$ for $i=1,2$ indicates the real and imaginary components of
the scalar field.
The second-order nonlinear wave equation becomes a system of three first-order
equations
\begin{eqnarray}
\dot{\hat{\phi_i}} & = &          \hat \Pi_i\,, \\
\dot{\hat{\Phi_i}} & = & \frac{1}{\cos^{d-2}x} \left( \cos^{d-1}x \,       \hat \Pi_i \right)_{,x}\,,\label{eq:eom_phi}\\
\dot{\hat{\Pi_i}}  & = & \! \! \frac{1}{\sin^{d-1}x}\! \left( \frac{\sin^{d-1}x}{\cos x}
                                                       \hat  \Phi_i
                                               \right)_{,x}
       \! \! \! + \! \left( |\hat \phi| \cos^{d-1}x \right)^{p-1} \hat \phi_i.
\label{kg}
\end{eqnarray}
The $U(1)$ symmetry of the complex scalar field results in a conserved
charge given by 
\begin{equation}
Q=\int_0^{\pi/2} dx\ \tan^{d-1} x \cos^{2\left(d-1\right)} x
                        \ \left(\hat \Pi_1 \hat \phi_2-\hat \Pi_2\hat \phi_1\right)  .
\label{eq:qinit}
\end{equation}
The total energy $E$ in the system is computed as
\begin{equation}
E=\int_0^{\pi/2} dx\ \tan^{d-1} x\ \cos^{2(d-1)}x
\left[   \frac{\hat \Phi_i^2}{\cos^2 x}+\hat \Pi_i^2
        +\frac{|\hat \phi|^{p+1}}{p+1}
\right],
\end{equation}
where sum over $i$ is implied.
The rescaled scalar quantities are subject to the boundary conditions
%\begin{eqnarray}
%\hat \phi_i(t,\pi/2) & = & 0 \\
%\hat \Phi_i(t,\pi/2) & = & 0 \\
%\hat \Pi_i(t,\pi/2) & = & 0.
%\end{eqnarray}
$\hat \phi_i(\pi/2,t)  =  0$, 
$\hat \Phi_i(\pi/2,t)  =  0$, and
$\hat \Pi_i(\pi/2,t)   =  0$.

The initial data used in this work is generalized from that presented
in~\cite{Bizon:2011gg} and is a subset of that used in~\cite{Buchel:2012uh}
\begin{eqnarray}
\hat \phi_i(x,0) & = & \frac{2\epsilon}{     \pi}
                   e^{-\frac{4\tan^2x}{     \pi^2\sigma^2}}\ \cos^{1-d}x\ \delta_i^1\\
\hat \Pi_i(x,0)  & = & \omega \hat \phi_j(x,0)\delta^1_j \delta^2_i,
\label{eq:id}
\end{eqnarray}
where $\sigma$, $\epsilon$, and $\omega$ are arbitrary constants. 
%where $\sigma=1/16$, and $\epsilon$ and $\omega$ are arbitrary constants. 
The initial  profile for $\hat \Phi$ is determined by the spatial derivative of the initial scalar field
\begin{equation}
\hat \Phi_i(x,0) = \left( 1-d \right) \tan^{d-2}x \, \hat \phi_i + \cos^{d-1}x \,  \hat \phi_{i,x}.
\label{eq:constraint}
\end{equation}
For $\omega \ne 0$, the initial data is charged.

As discussed in~\cite{Liebling:2005ap,2004Nonli..17.2187B}, when the amplitude
of the scalar field becomes large, the nonlinear potential term dominates and focuses 
the scalar pulse, leading to singularity formation at some collapse time $t_c$. Numerically, such
collapse is ``detected'' by stopping the code when 
$|\phi| \ge 10$, where the value $10$ is arbitrary.

These equations are solved using the same infrastructure and methods 
described in~\cite{Bizon:2011gg}, although 
%no metric terms are required
no dynamical evolution of metric variables is required as AdS is treated
as a fixed background.
%since the model evolves in pure AdS.
Tests of the code indicate that
it converges to better than third order in the grid spacing and conserves
total charge and energy. The convergence order and fractional changes
in energy and charge are presented for a typical run in Fig.~\ref{fig:convergence}.\\

\begin{figure}
\vspace{-0.75in}
\centerline{\includegraphics[width=4in]{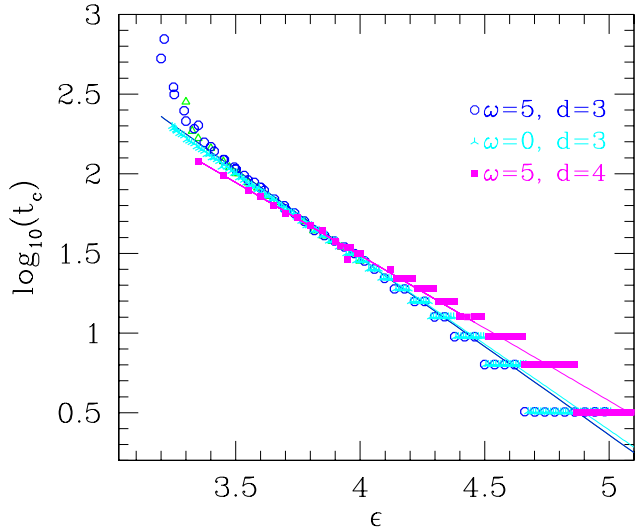}}
\vspace{-1.15in}
\caption{Time to collapse as a function of $\epsilon$ in the initial data. As
$\epsilon$ decreases, the initial pulse executes more bounces and the collapse time
increases. The jumps in the collapse times shown correspond to the initial pulse executing
another bounce and therefore
collapse occurs roughly in multiples of $\Delta t = \pi$.
Collapse times roughly follow $\log_{10} (t_c) = A \epsilon + B$.
Compare to the bottom frame of Fig. 1 of~\cite{Jalmuzna:2011qw} which shows the time of
collapse to black hole for a scalar field.
Charged (blue, open circles) initial data follows a trendline with $A=-1.11$ and $B=5.91$, 
while uncharged (cyan stars) configurations follow $A=-1.08$ and $B=5.78$.
A charged configuration
in $d=4$ (magenta solid squares) follows a slightly less negative trend with
$A=-0.914$ and $B=5.14$.
A few higher resolution results for $\omega=5$, $d=3$ are also shown (open, green triangles),
appearing largely consistent with the normal resolution data.
}
\label{fig:collapsetimes1}
\end{figure}

\noindent \emph{\bf Results:}
Typical results are represented in Fig.~\ref{fig:collapsetimes1}.
Charged (blue open circles), uncharged (cyan stars), and charged higher
dimensional (magenta, solid squares) evolutions are shown for a wide
range of $\epsilon$. The collapse times get progressively longer (roughly in multiples of
the crossing-time $\pi$). The important point here is that the scalar executes
an increasing number of reflections off the boundary with
decreasing $\epsilon$ and eventually collapses.

As also found in~\cite{Buchel:2012uh} for the gravitating scalar, the
global charge does not appear to have any significant effect.
A careful reader might observe, however, that collapse times of the charged
family in $d=3$ rises quite sharply at small $\epsilon$. For such long runtimes,
higher resolution is generally called for, despite the use of AMR because of
the accumulation of error, and higher resolution results are shown in green,
open triangles. The two resolutions mostly agree with each other, and so perhaps
this rapid rise in collapse times is evidence of some physical effect of either the charge
or other difference in the initial data. An example of one such property of initial
data resulting in the rapid rise in collapse times can be found in~\cite{Buchelboson}, but
further calculations are needed for this case.

% Cured by higher resolution:
%
%The behavior of the $d=4$ runs for small $\epsilon$ is quite different than
%the larger $\epsilon$ runs.
 %They nevertheless still collapse. This differing behaviour is
%likely due to higher numerical noise
%in higher dimensions for the longest runs, but it is still being investigated.

\begin{figure}
\vspace{-0.75in}
\centerline{\includegraphics[width=4in]{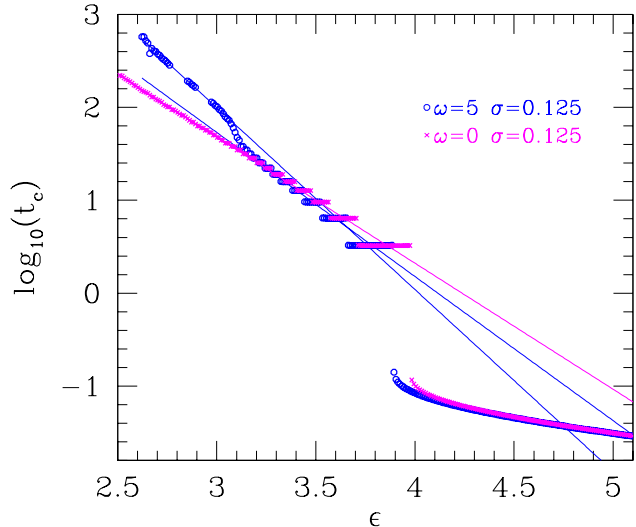}}
\vspace{-1.15in}
\caption{Collapse times for families characterized by $\sigma=0.125$ (in
contrast to families with $\sigma=0.0625$ shown in
Fig.~\ref{fig:collapsetimes1}).
Initial data with rotation~(blue circles) displays a transition from $A=-1.55$ and $B=6.38$ to $A=-1.96$ and $B=7.87$.
Initial data without rotation~(magenta crosses) follows the trendline described by
$A=-1.36$ and $B=5.75$.
}
\label{fig:collapsetimes3}
\end{figure}

The collapse times for a different family of initial data are shown
in Fig.~\ref{fig:collapsetimes3}. This family is characterized by $\sigma=1/8$,
twice the value characterizing the data shown in Fig.~\ref{fig:collapsetimes1}.
These data similarly suggest scalar collapse is inevitable for any initial
amplitude.

%%%%%%%%%%%%%%%%%%%%%%%%%%%%%%%%%%%%%%%%%%%%%%
%%%%%%%%%%%%%%%%%%%%%%%%%%%%%%%%%%%%%%%%%%%%%%
%%%%%%%%%%%%%%%%%%%%%%%%%%%%%%%%%%%%%%%%%%%%%%
%%%%%%%%%%%%%%%%%%%%%%%%%%%%%%%%%%%%%%%%%%%%%%
%%%%%%%%%%%%%%%%%%%%%%%%%%%%%%%%%%%%%%%%%%%%%%
\begin{figure}
\vspace{-0.75in}
\centerline{\includegraphics[width=4in]{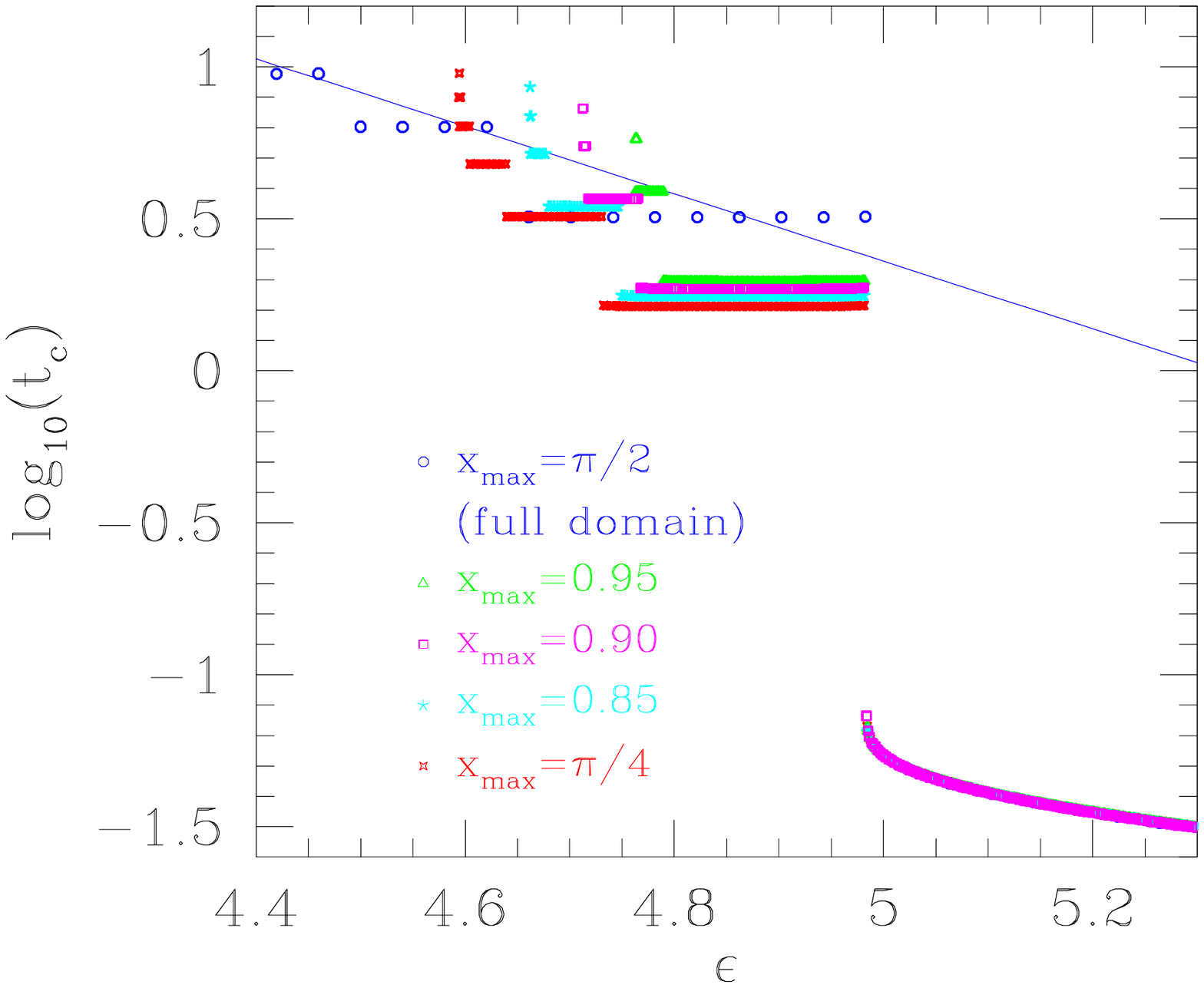}}
\vspace{-2.4in}
\centerline{\includegraphics[width=4in]{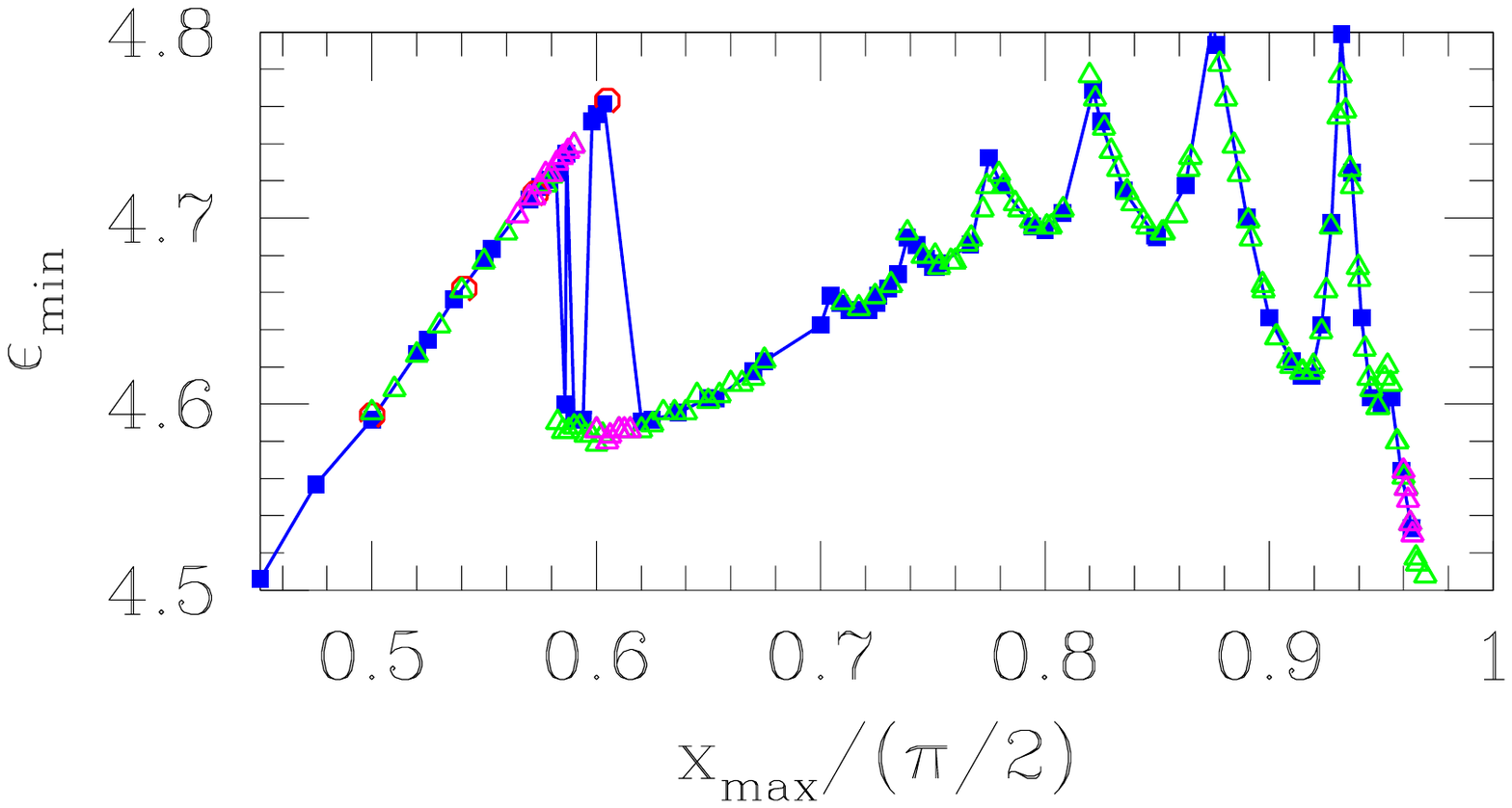}}
\vspace{-2.45in}
\caption{Collapse within restricted domains of AdS. {\bf Top:} Collapse
times for various domains extending to $x_{\rm max}$. The same results
for the $\omega=5$, $d=3$ evolutions in the full domain from Fig.~\ref{fig:collapsetimes1} are shown to facilitate comparison and contrast between full-domain
and finite-domain results. Note that as the outer boundary is moved inwards,
the minimum value of $\epsilon$ that produces collapse decreases.
For 
sufficiently small $\epsilon$, no collapse is detected and therefore collapse
times for these runs extend mostly vertically.
The collapsing solutions closest to the threshold have values of $\epsilon$
within a few parts in a million of the $\epsilon$ values of
solutions which do not collapse.
{\bf Bottom:} An explicit calculation of the minimum $\epsilon_{\rm min}$ for
which collapse is detected as a function of $x_{\rm max}$. To get a sense of
how converged these values (blue, solid squares) are, higher resolution computations (open, green triangles) and higher
resolution computations to later times ($t_{\rm max}=50$) are also shown. Finally, the $\epsilon_{\rm min}$ values (open, red circles)
 from the data shown in the top frame are also displayed. The apparent linearity of the top frame belies the various features in
the full ``spectrum."
%are the values corresponding to the results in the top frame from which one can
%see that they occur in a linear region.
}
\label{fig:collapsetimes2}
\end{figure}
%%%%%%%%%%%%%%%%%%%%%%%%%%%%%%%%%%%%%%%%%%%%%%

Ref.~\cite{Buchel:2012uh} studied gravitational collapse on a restricted
computational domain by imposing reflecting boundary conditions at some
largest value $x_{\rm max}$ of the coordinate $x$. In such a space, the
scalar pulse does not ``see'' the full AdS background but nevertheless 
still evolves within a bounded domain. That the evolutions showed collapse
after a number of reflections supported the view that the cause behind
the inevitable collapse was that the domain was bounded, not that the space
was AdS in particular. The boundedness of the domain allowed for the
nonlinearity of gravity to act continuously. However, Ref.~\cite{Buchel:2012uh}
also reported that, although collapse did occur after a number of reflections,
for small enough initial amplitude, collapse did \emph{not} occur, even
after many reflections.

In light of this interesting effect, evolutions of the nonlinear wave equation
here are conducted within a restricted domain with the identical, reflecting
boundary condition of Ref~\cite{Buchel:2012uh}.
In particular, the scalar field is fixed via enforcement of a Dirichlet condition $\phi_i(x_{\rm max},t)=0$ and $\Pi_i(x_{\rm max},t)=0$.
Studies with domains extending only to $x_{\rm max}
= 0.95, 0.90, 0.85,$ and $\pi/4 \approx 0.785$ are shown in the top frame of Fig.~\ref{fig:collapsetimes2},
along with those within the full AdS spacetime~($x_{\rm max} = \pi/2$) for comparison.

These restricted domain runs behave in
stark contrast with the full-domain results. Evolutions which collapse essentially
immediately (before any reflection off the boundary) nearly coincide as would be expected.
After one reflection, one would expect some range of $\epsilon$-values that would result
in collapse, and therefore these restricted domains collapse faster than the full-domain
evolutions simply because the geometric ``concentration'' at the origin happens earlier.
However, as one considers even smaller $\epsilon$ values, one sees that the collapse time
grows apparently asymptotically to infinity. In other words, the evolutions suggest the existence 
of some
threshold value of $\epsilon_{\rm min}$ below which finite-domain evolutions no longer collapse. 

As $\epsilon$ is decreased in increasingly finer steps near the threshold, 
more bounces are observed before collapse, as indicated in the top frame of Fig.~\ref{fig:collapsetimes2}. It
is not clear yet what happens in the limit that $\epsilon$ approaches some critical $\epsilon_{\rm min}$,
and in particular whether the number of bounces continues to increase.

In the examples shown in the top frame of Fig.~\ref{fig:collapsetimes2}, this minimum
value increases apparently linearly with increasing $x_{\rm max}$. However,
the results of a brief study of $\epsilon_{\rm min}$
as a function of $x_{\rm max}$ is shown in the bottom frame and reveals very nontrivial behavior.
These values were obtained by a bisection search on $\epsilon$ for a given value of $x_{\rm max}$. Beginning
with a bounding bracket $\left[\epsilon_{\rm low},\epsilon_{\rm high}\right]$, the evolution of the
average $\epsilon_{\rm avg}=\left( 1/2\right) \left[ \epsilon_{\rm low}+\epsilon_{\rm high}\right]$ was
computed up to a maximum time $t_{\rm max}=30$. A new bracket is found such that if collapse occurred then one resets $\epsilon_{\rm high}=\epsilon_{\rm avg}$, and otherwise one sets $\epsilon_{\rm low}=\epsilon_{\rm avg}$.
Most of the computing time is spent on evolutions which do not collapse and therefore limiting $t_{\rm max}$ greatly
speeds-up the calculation at the expense of some accuracy. The search is carried out until the fractional difference
in the bounds is one-ten-thousandth.

It is interesting to note that because the full domain is expected to collapse for any value
of $\epsilon$, its respective $\epsilon_{\rm min}$ is zero. In addition,  the results shown in the
top frame of Fig.~\ref{fig:collapsetimes2}  indicate an increasing $\epsilon_{\rm min}$ with
increasing $x_{\rm max}$. However, as shown in the bottom frame, at large values of $x_{\rm max}$ this
behavior reverses giving one hope that the limit $x_{\rm max}\rightarrow \pi/2$ is smooth.
Higher resolution data and data with a larger value of $t_{\rm max}$ are also shown and appear to support
the validity of these surprisingly intricate data.
 Further calculations
are needed especially in the apparently oscillatory region around $x_{\rm max}\approx 0.6$.
As noted below, it would also be interesting to compare this behavior with results using a Neumann boundary condition
%at some $x_{\rm max}$.
instead.

That small-$\epsilon$, restricted-domain evolutions fail to collapse argues against
the idea that the important effect of AdS is its introduction of reflections. Instead
perhaps some other property of AdS accounts for the inevitable collapse and by
restricting the domain, this property is affected. That is not the argument presented
here. Instead, there appears to be another effect introduced by the
restriction of the domain, an effect not seen in the full domain case.

\begin{figure}
\vspace{-0.75in}
\centerline{\includegraphics[width=4in]{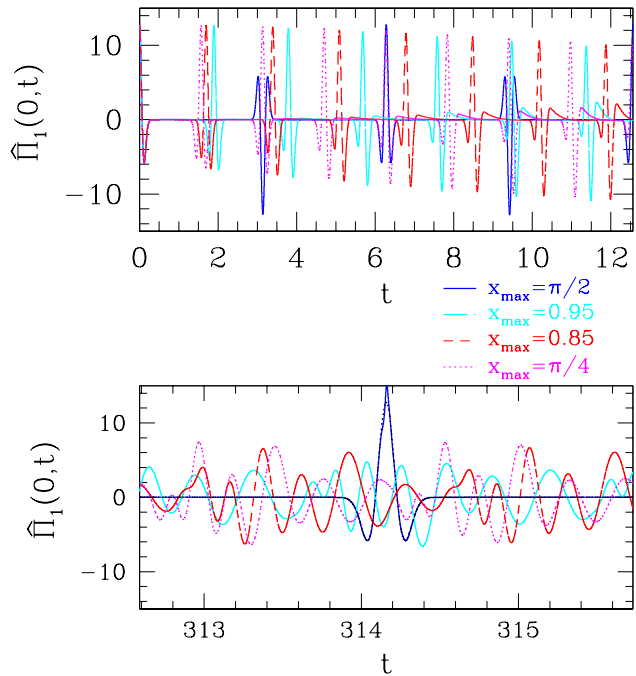}}
\vspace{-1.15in}
\caption{Demonstration of dispersion with restricted domain evolutions of
the ``linear'' wave equation. Shown is the behavior of $\hat \Pi_1$ at the origin
versus time for evolutions in which the nonlinear term $\phi^7$ has been
removed. For reference is shown the full domain result (solid blue) which
maintains its shape (except for an inversion of sign) with each reflection.
The restricted domain evolutions show no inversion but do change shape.
The bottom frame shows the late time behavior where the dispersion of the
restricted domain evolutions is quite apparent. Also shown (dotted black) is the
full domain pulse from $t\approx \pi$, shifted in time and inverted in sign. That
it nearly overlays the late-time, full domain result indicates a lack of
dispersion when evolving in the full AdS space.
}
\label{fig:dispersion}
\end{figure}

Consider the wave equation with no nonlinear term, that is Eq.~(\ref{kg}) without
the $\phi^{p-1}$ term\footnote{Qualitatively identical results are obtained in the appropriately weak field regime of the full, nonlinear equation.}. In spherically symmetric Minkowski space, the solutions take
the simple form $r \phi(r,t) = f(r\pm t)$ where $r$ is the radial coordinate and $f()$
is some function. There is no dispersion of $f()$, although $\phi(r,t)$ decreases
in amplitude with increasing $r$. In AdS, however, things are not so simple and one expects that
the metric terms will cause dispersion. In Fig.~\ref{fig:dispersion}, the behavior
of $\hat \Pi_1(0,t)$ at the origin is shown at various times. In the full
domain, the pulse ``reflects'' off the AdS boundary without inversion and
flips when it implodes through the origin. Besides the inversion, the behavior
of the pulse remains the same, even at late times.
The periodicity results from the fact that the modes of
the scalar field in AdS have integer values, and hence a generic initial
configuration will repeat with a period (at most) of $\Delta t=2\pi$~\endnote{Thanks to
Carsten Gundlach for pointing this out.}.
In the restricted domains,
the pulse inverts at both the outer boundary\footnote{One could instead
impose a free Neumann boundary condition at some $x_{\rm max}$ that would
not invert the pulse at the outer boundary and thus be more similar to the
AdS boundary. However, initial attempts at such a condition proved (numerically) unstable.}
 and the origin, and it disperses.
Because the restricted domain does not allow for integer eigenvalues, it is
not periodic.

The results suggest that this dispersion competes with the nonlinearity (be it
gravitational or simply some scalar potential term). In this way, restricted
domain evolutions of small initial data are dominated by the dispersion whereas
strong initial data are instead dominated by the nonlinearity. In the
special case where the full domain is allowed, no dispersion occurs, and any
initial data will eventually collapse.\\

%%%%%%%%%%%%%%%%%%%%%%%%%%%%%%%%%%%%%
%%%%%%%%%%%%%%%%%%%%%%%%%%%%%%%%%%%%%
%%%%%%%%%%%%%%%%%%%%%%%%%%%%%%%%%%%%%

\noindent \emph{\bf Conclusions:}
%These studies suggest that the bounded domain provided by AdS can result 
These studies provide an example of a non-gravitational, hyperbolic
system which results in sharpening, and eventual collapse, of an initial pulse.
This result appears analogous to the gravitational collapse of a scalar field in AdS.

Evolutions within a restricted domain with reflecting boundary conditions
also behave similar to the gravitating scalar~\cite{Buchel:2012uh}.
In particular, such evolutions
collapse for only a limited range of decreasing amplitude and stop collapsing below
some threshold initial size.
The imposition of a reflecting boundary condition introduces dispersion where it
would otherwise not appear, and this dispersion appears to compete with the
weak turbulence~\cite{Bizon:2011gg} that transfers energy to higher frequencies. 
As one considers smaller initial data, one observes a transition from the dominance
of weak turbulence to dispersion.
%They also
%indicate an interesting effect on restricted domains in which inevitable collapse
%only occurs down to some limiting value strength parameter, an effect also seen
%in the gravitating case~\cite{Buchel:2012uh}.

It should be noted that not all scalar configurations in the gravitating case
are unstable~\cite{Buchelboson,Dias:2012tq},
and it would be interesting to study the $p=5$ case of the semilinear equation because, at least in Minkowski space, it possesses a static solution~\cite{Liebling:2005ap,2004Nonli..17.2187B,szpak}.

%Remaining questions:
%\begin{itemize}
%\item{Does the inevitable collapse in the full domain continue as $\epsilon \rightarrow \infty$?}
%\item{In Bizon's work $t_{\rm AH} \propto \epsilon^{-2}$, but here $t_{\rm c} \propto e^{-\epsilon}$...why? Can the rate be estimated by perturbative work? Is the rate somehow universal for different families? Different for distinct dimension $d$?}
%\item{What does the apparent transition in the $\sigma=0.125$ family from one linear regime to the other indicate?}
%\item{What does the cut-off in collapse at small $\epsilon$ of finite-domain runs indicate? How does it depend on $x_{\rm max}$? Is there interesting behavior in the continuum model at this threshold?}
%\end{itemize}
%Future: other values of $p$, perturbative studies of the semilinear wave equation

~\\
\noindent{\bf{\em Acknowledgments:}}
Thanks go to Alex Buchel, Carsten Gundlach, Luis Lehner, Oscar Reula,
and Nikodem Szpak for helpful
discussions.
%It is a pleasure to thank 
%for interesting and helpful discussions.
This work was supported by the NSF via grant PHY-0969827.
%Research at Perimeter
%Institute is supported through Industry Canada and by the Province of Ontario
%through the Ministry of Research \& Innovation.
Computations were performed thanks to allocations at the Extreme Science and Engineering Discovery Environment~(XSEDE),
which is supported by National Science Foundation grant number OCI-1053575.

%%%%%%%%%%%%%%%%%%%%%%%%%%%%%%%%%%%%%%%%%%%%%%%%%%%%%%%%%%%%%%%%%%%%
%
%   B I B L I O G R A P H Y
%
%%%%%%%%%%%%%%%%%%%%%%%%%%%%%%%%%%%%%%%%%%%%%%%%%%%%%%%%%%%%%%%%%%%%
%\bibliographystyle{apsrev}
%\bibliographystyle{ieeetr}
\bibliographystyle{utphys}
\bibliography{./the}

\providecommand{\href}[2]{#2}\begingroup\raggedright\begin{thebibliography}{1}

\bibitem{Bizon:2011gg}
P.~Bizo\'n and A.~Rostworowski, ``{On weakly turbulent instability of anti-de
  Sitter space},'' \href{http://dx.doi.org/10.1103/PhysRevLett.107.031102}{{\em
  Phys.Rev.Lett.} {\bfseries 107} (2011) 031102},
\href{http://arxiv.org/abs/1104.3702}{{\ttfamily arXiv:1104.3702 [gr-qc]}}.
%%CITATION = ARXIV:1104.3702;%%.

\bibitem{Jalmuzna:2011qw}
J.~Jalmuzna, A.~Rostworowski, and P.~Bizo\'n, ``{A Comment on AdS collapse of a
  scalar field in higher dimensions},''
  \href{http://dx.doi.org/10.1103/PhysRevD.84.085021}{{\em Phys.Rev.}
  {\bfseries D84} (2011) 085021},
\href{http://arxiv.org/abs/1108.4539}{{\ttfamily arXiv:1108.4539 [gr-qc]}}.
%%CITATION = ARXIV:1108.4539;%%.

\bibitem{Liebling:2005ap}
S.~L. Liebling, ``{Threshold of singularity formation in the semilinear wave
  equation},'' \href{http://dx.doi.org/10.1103/PhysRevD.71.044019}{{\em
  Phys.Rev.} {\bfseries D71} (2005) 044019},
\href{http://arxiv.org/abs/gr-qc/0502056}{{\ttfamily arXiv:gr-qc/0502056
  [gr-qc]}}.
%%CITATION = GR-QC/0502056;%%.

\bibitem{2004Nonli..17.2187B}
P.~{Bizo\'n}, T.~{Chmaj}, and Z.~{Tabor}, ``{On blowup for semilinear wave
  equations with a focusing nonlinearity},''
  \href{http://dx.doi.org/10.1088/0951-7715/17/6/009}{{\em Nonlinearity}
  {\bfseries 17} (Nov., 2004) 2187--2201},
  \href{http://arxiv.org/abs/arXiv:math-ph/0311019}{{\ttfamily
  arXiv:math-ph/0311019}}.

\bibitem{Buchel:2012uh}
A.~Buchel, L.~Lehner, and S.~L. Liebling, ``{Scalar Collapse in AdS},''
  \href{http://dx.doi.org/10.1103/PhysRevD.86.123011}{{\em Phys. Rev. D}
  {\bfseries 86} (Dec, 2012) 123011},
\href{http://arxiv.org/abs/1210.0890}{{\ttfamily arXiv:1210.0890 [gr-qc]}}.
%%CITATION = ARXIV:1210.0890;%%.

\bibitem{Buchelboson}
A.~Buchel, S.~L. Liebling, and L.~Lehner, ``{Boson stars in AdS},'' {\em in
  preparation} (2012) .

\bibitem{Dias:2012tq}
O.~J. Dias, G.~T. Horowitz, D.~Marolf, and J.~E. Santos, ``{On the Nonlinear
  Stability of Asymptotically Anti-de Sitter Solutions},''
  \href{http://dx.doi.org/10.1088/0264-9381/29/23/235019}{{\em
  Class.Quant.Grav.} {\bfseries 29} (2012) 235019},
\href{http://arxiv.org/abs/1208.5772}{{\ttfamily arXiv:1208.5772 [gr-qc]}}.
%%CITATION = ARXIV:1208.5772;%%.

\bibitem{szpak}
N.~Szpak, ``{Relaxation to Intermediate Attractors in Nonlinear Wave
  Equations},'' \href{http://dx.doi.org/10.1023/A:1010460004007}{{\em
  Theoretical and Mathematical Physics} {\bfseries 127} (2001) 817--826}.

\end{thebibliography}\endgroup

\end{document}